\documentclass[prl,twocolumn,letterpaper,showpacs,superscriptaddress]{revtex4}
\usepackage{graphicx,bm,amssymb,amsmath}

\newcommand{\DD}{\text{D}}
\newcommand{\dd}{\text{d}}

\def\be{\begin{equation}}
\def\fin{\end{equation}}
\def\disp{\displaystyle}

\def\T{{\sf T\kern-.45em T}}
\def\C{\kern.1em{\raise.47ex\hbox{$\scriptscriptstyle |$}}
       \kern-.40em{\sf C}}

\begin{document}
\title{Pushing off the walls: a mechanism of cell motility in confinement}
\date{5 February 2009}
\author{ R.J. Hawkins}
\affiliation{UMR 7600, Universit\'e Pierre et Marie Curie/CNRS, 4 Place Jussieu, 75255
Paris Cedex 05 France}
\author{M. Piel}
\affiliation{UMR 144, Institut Curie/CNRS, 
26 rue d'Ulm 75248 Paris Cedex 05 France}
\author{G. Faure-Andre}
\affiliation{U 653, Inserm/Institut Curie, 
26 rue d'Ulm 75248 Paris Cedex 05 France}
\author{A.M. Lennon-Dumenil}
\affiliation{U 653, Inserm/Institut Curie, 
26 rue d'Ulm 75248 Paris Cedex 05 France}\author{J.F. Joanny}
\affiliation{UMR 168, Institut Curie/CNRS, 
26 rue d'Ulm 75248 Paris Cedex 05 France}
\author{ J. Prost}
\affiliation{UMR 168, Institut Curie/CNRS, 
26 rue d'Ulm 75248 Paris Cedex 05 France}
\affiliation{ E.S.P.C.I, 10 rue Vauquelin, 75231 Paris Cedex 05, France
}
\author{ R.Voituriez}
\affiliation{UMR 7600, Universit\'e Pierre et Marie Curie/CNRS, 4 Place Jussieu, 75255
Paris Cedex 05 France}


\begin{abstract}
 We propose a novel, mechanism of cell motility, which relies on the coupling of actin polymerization 
at the cell membrane to geometric confinement. We consider a polymerizing viscoelastic cytoskeletal gel 
confined in a narrow channel, and show analytically that spontaneous motion occurs. Interestingly, this 
does not require specific adhesion with the channel walls, and yields velocities potentially larger than 
the polymerization velocity. The contractile activity of myosin motors is not necessary to trigger 
motility in this mechanism, but is shown quantitatively to increase the velocity. Our model qualitatively 
accounts for recent experiments which show that cells without specific adhesion proteins are motile only 
in confined environments while they are unable to move on a flat surface, and could help in understanding 
the mechanisms of cell migration in more complex confined geometries such as living tissues.
\end{abstract}
\maketitle

Besides its obvious importance for biology, cell motility has
motivated numerous studies in the physics community. Identifying
simple mechanisms of self-motion of soft condensed matter is an
important challenge for physics, cell biology and biomimetic material
technology. Sustained motion at low reynolds number
necessitates a constant energy input, and
therefore an active system, that is a system driven out of equilibrium
by an internal or an external energy source.   The cell cytoskeleton
is a striking example of such an active system. It is a network of long
semi-flexible filaments made up of protein subunits, interacting with
other proteins 
such as 
motor proteins
which
use 
the 
chemical energy 
of ATP hydrolysis 
to  exert active stresses that
deform the network \cite{Nedelec1997}.
Other examples of active systems range from
animal flocks to 
bacterial colonies~\cite{dombrowski:098103,Hatwalne2004} and vibrated granular
media \cite{Narayan:2007tw}.

Modelling cell motility  has inspired both experimentalists and
theoreticians, who have now distinguished two main different
mechanisms: polymerization (treadmilling) and contractility. Both polymerization induced motion \cite{Noireaux:2000ia,Bernheim-Groswasser:2002xa} and contractility induced spontaneous
flows \cite{Nedelec1997}  have now been observed in-vitro, and studied
theoretically \cite{Kruse+,Voituriez2005+6,zum2008}
and numerically \cite{Marenduzzo:2007+8}. In all models the key ingredients of motion are an energy input
to compensate dissipation and sufficient adhesion or
friction with a substrate to acquire momentum. The usual
picture of cell locomotion is then as follows: the cell lamellipodium builds strong adhesion
points with the substrate and pushes forward its membrane by
polymerizing actin. At the back, the cell body contracts and breaks
the adhesion points. In particular, the overall cell velocity is then
limited by the actin polymerization rate (which however varies substantially between cell types), in agreement with 
available experimental data  \cite{Theriot1991,Clainche2008,Pollard2003}.

\begin{figure}[hbt]
\begin{centering}
\includegraphics[width=0.5\textwidth]{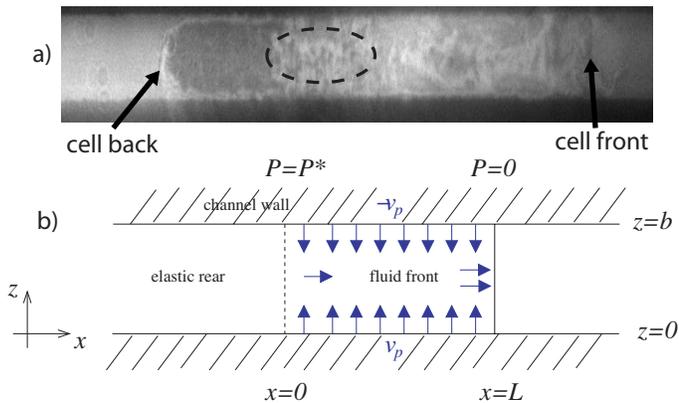}
\caption{\label{fig_geom}a) RICM image of a dendritic cell moving (to the right) in a channel of $4 \mu\text{m}$ width. The dark zone at the back of the cell (left) indicates a large contact of the membrane with the channel wall (independent of the nucleus, dotted line), and therefore a high normal constraint, compared to the front (right). The typical observed velocity reaches  $12-15\mu\text{m/min}$ in channels and  $4-6\mu\text{m/min}$ on a flat surface. b) Channel geometry and model. The arrows show the flow direction.}
\end{centering}
\end{figure}

In a recent paper \cite{Lammermann:2008tf} (see also \cite{Malawista1997} for another cell type), it has been observed both
in-vivo and in-vitro that 
mutant dendritic cells (DC) that are unable to produce active integrin complexes (adhesion proteins) display sustained motility in confined
environments (tissues or synthetic polymeric gels), whereas they fail
to move on flat two-dimensional substrates due to their reduced
adhesion ability. These observations suggest the existence of an alternative mechanism of motility to the adhesion dependent picture outlined above. Here we propose a new, simple mechanism of motility which accounts for  these observations. This mechanism  is mainly powered by actin  polymerization  at the cell membrane,  and strongly relies on geometric
confinement.  Interestingly, it does not necessitate strong specific
adhesion,
and yields velocities potentially larger than the
polymerization velocity.  This confinement induced motility mechanism is backed  by
in-vitro experiments of DC motility in microfabricated channels \cite{note} (see Fig. \ref{fig_geom}). 

We first introduce our  model in its minimal form
of a polymerizing  visco-elastic gel confined
in a channel. 
We then refine this model to mimic motile
cells in confinement.
Finally, we show quantitatively that the contractile activity of myosin increases the velocity of motion.

The model, which relies on the hydrodynamic theory of active gels
\cite{Kruse+}, is as follows.  We consider an
incompressible viscoelastic film confined in a  bidimensional channel of width $b$. Note that this bidimensional geometry mimics the experimental conditions of channels of rectangular section (Fig. \ref{fig_geom}), and that the case of a cylindrical confining channel can be treated with minor modifications. 
The axes are defined with $x$ along the
channel and $z$ across it. The confining walls are placed at $z=0$ and
$z=b$.
We denote the components of
the velocity of the gel by $v_i$, and the strain rate by $u_{ij}=(\partial_i v_j+\partial_j
v_i)/2$. We assume that the gel is described by a
linear Maxwell model of visco-elasticity.  The constitutive equation
relating the strain rate to the deviatory stress tensor $\sigma_{ij}$ is then
written $ 2\eta
u_{ij}=\left(1+\tau\frac{\DD}{\DD t}\right)\sigma_{ij},$ where $\eta$ is the shear viscosity, $\tau$ is a
typical relaxation time and $\DD/\DD t$ denotes the convective derivative. For dilute polymer gels, $\tau$ is very small
and the gel behaves as a viscous fluid. At higher concentrations, or
for more cross-linked gels, $\tau $ becomes very large and the gel
behaves as an elastic medium.  In what
follows we assume that the gel is in either of the two regimes. For
the sake of simplicity we assume here that the gel is incompressible,
which means that $\tau$ depends on pressure $P$ only. We define a
critical pressure such that $\tau(P<P^*)=0$ (viscous regime) and
$\tau(P>P^*)=\infty$ (elastic regime).  
We next suppose  that the gel is polymerized at the gel/substrate interface
with speed $v_p\equiv v_z(x,z=0)=-v_z(x,z=b)$ in the viscous regime, as depicted in Fig.~\ref{fig_geom}.
This is justified by the common observation 
of actin polymerization activators such as WASP proteins preferentially
located along the cell membrane \cite{alberts,Clainche2008}. In the case of DCs,  actin filaments can anchor
perpendicularly to the cell membrane, forming structures called
podosomes where polymerization takes place, therefore inducing an
inward flow of actin as it is assumed in our model \cite{Calle2006}.
Finally we assume viscous friction at the channel walls $z=0$ and $z=b$, and write $\sigma_{xz}= \xi v_x$ where $\xi$ characterizes the friction.

We now derive the dynamical equations of the system in the
lubrication approximation  ( $b\ll L$ where $L$ is the typical
length of the system). 
In this limit the Reynolds
number is small and the velocity field $v_i(x,z)$ can be obtained from the force balance 
$\partial_x P=\eta\partial_z^2 v_x$ and 
the condition of  incompressibility 
$u_{xx}+u_{zz}=0$. Defining the average velocity along the
channel $v(x)=(1/b)\int_0^b v_x(x,z)dz$, we obtain the following  Darcy's law:
\begin{equation}
\label{darcy}
v(x)=-\frac{b^2}{12\eta}\left(1+\frac{6\eta}{b}\xi^{-1}\right)\frac{\dd P}{\dd x}.
\end{equation} 
Including the depolymerization of the gel $k_d$, mass conservation of the gel reads
$\frac{\dd v}{\dd x}=2v_p/b-k_d$,
which gives in turn:
\begin{equation}
\label{Peq}
\frac{\dd}{\dd x}\left[(1+{\tilde \xi}^{-1}) \frac{\dd P}{\dd x}\right] =\frac{12\eta (2 v_p -bk_d)}{b^3},
\end{equation} where  $v_p$ and  the nondimensional friction  ${\tilde \xi}\equiv\xi b/6\eta$  can be a priori
functions of $P$ and $x$.

Two boundary conditions are
needed to determine the pressure profile $P(x)$. We
neglect the friction with the surrounding fluid in the channel and
set the pressure at the leading edge, which is assumed to coincide with the point $x=L$, as $P(L)=0$, which gives the first
boundary condition. Note that if the pressure at the leading edge is finite due to an external force, our results apply with  an unimportant shift in the pressure field. 
We look for stationary states with broken symmetry and positive velocity and therefore the
pressure is a decreasing function of $x$. We then argue
that if the system is large enough,  there exists a travelling front
of gel of length $L$
in the fluid phase,
travelling at velocity $V$. The back boundary of this front fluid part
coincides with the point  $x=0$ where the pressure reaches the threshold $P^*$,
behind which is a growing elastic part. 
Such a denser elastic region at the back of DCs, called the uropod, is indeed well reported \cite{Serrador1999}, and is characterized  by a higher concentration of  cross-linkers. As the velocity of the elastic part
should be zero one has $v(0)=0=\frac{\dd P}{\dd x}|_{x=0}$, giving the
second boundary condition which allows the explicit calculation of the
pressure field. The self-consistent condition $P(0)=P^*$ gives in turn
an equation enabling the calculation of the length $L$ of the fluid
front. We then  write that the velocity $V$ of the front is given by
the calculated velocity of the flow plus the polymerization velocity at the leading edge $v(L)+v_p(L)$. We stress that the flow velocity is {\it forward}, i.e. in the same direction as the moving leading edge. Note that the length $L$ of
the fluid front is constant, which necessitates that the elastic/fluid boundary moves at the same velocity $V$.

Qualitatively, the value of the length $L$ is dictated by the
steepness of the pressure gradient, and therefore by the friction
$\tilde \xi$. If $\tilde\xi$ is small, then only very long fluid fronts can
move. We show now quantitatively that the coupling of $\tilde\xi$ with the
pressure field actually enables short fluid fronts to move even with a
low bare friction. The key ingredients are as follows. Following
\cite{FabienGerbal11012000} we argue that the friction coefficient
$\tilde\xi$  depends on the normal constraint in the case of a polymeric
gel. Indeed qualitatively a high normal constraint increases the
attachment rate of polymers onto the channel walls by lowering the
entropic barrier, and decreases the detatchment rate. It is
shown  in
\cite{FabienGerbal11012000} that  in the regime of moderate tangential
speed, one has 
$ {\tilde\xi}={\tilde\xi}_0 e^{\beta(P-\sigma_{nn})}$,
where in our geometry the normal stress $\sigma_{nn}=\sigma_{zz}$ for both walls at $z=0$ and $z=b$. 
Next, following standard ratchet models \cite{Dogterom2002}, the
polymerization speed at the cell membrane  is assumed to depend on the
normal constraint in the gel according to $v_p=v_p^0
e^{-\alpha(P-\sigma_{zz})}$. 
These assumptions make Eq.~(\ref{Peq}) an autonomous 
equation for $P$, which is completed by the two boundary conditions
discussed above. 

\begin{figure}[hbt]
\begin{centering}
\includegraphics[width=0.4\textwidth]{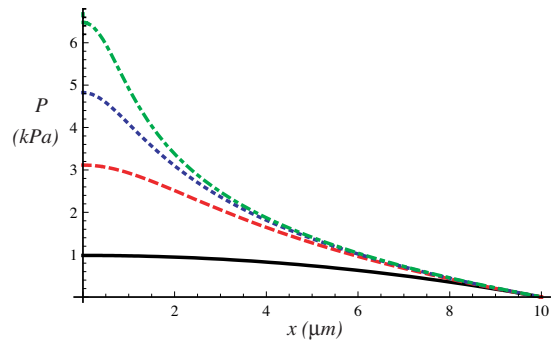}
\vspace{-1cm}
\caption{
\label{fig_Pprofile}Pressure profile. Dashed red: numerical value $P^0$ for $\beta=1$ and $\alpha=0.01\text{kPa}^{-1}$, dotted blue: analytical expression Eq.~(\ref{Pex}) for $\beta=1$, $\alpha=0$, dot-dashed green: numerical value $P^1$ for $\beta=1$, $\alpha=0$, solid black: numerical value $P^0$ for $\beta=0$, $\alpha=0.01\text{kPa}^{-1}$. 
Other parameters (estimates from \cite{FabienGerbal11012000,Callan-Jones:2008sf}): $L=10\mu\text{m}$, $b=1\mu\text{m}$, $\eta=10\text{kPas}$, $k_d=0.1s^{-1}$, $v_p^0=0.1\mu\text{m}s^{-1}$, $\xi_0=0.1\text{kPa s }\mu\text{m}^{-1}$. $v_p^0$ is taken as the speed of DCs on a surface which is expected to be the actin polymerisation speed \cite{Theriot1991}. $\xi_0$ is taken as very small (lowest estimate in \cite{FabienGerbal11012000} which is 100 fold smaller than in keratocytes \cite{Kruse+}) to mimic the low adhesion of integrin knockout DCs.}
\end{centering}
\end{figure}

To our knowledge, such an equation cannot be solved analytically in
the general case.  In the regime of small $\alpha$ (defined
by $\alpha P^*\ll 1$), $v_p$ can be taken as constant and  an analytical approximation scheme  can be proposed, which 
enables a discussion of the motility mechanism. We first
neglect the elongational shear stress 
and write ${\tilde \xi}\approx {\tilde \xi}_0 e^{\beta P}$. This
assumption underestimates the friction, and therefore the pressure
field.  Eq. \ref{Peq} can then be integrated and yields an
implicit equation for the pressure field to lowest order $P^0$:
\begin{equation}
\label{Pex} 
\disp P^0+\frac{{\tilde \xi_0}^{-1}}{ \beta}(1-e^{-\beta P^0})=\frac{6\eta (2v_p^0-bk_d)}{b^3}(L^2-x^2).
\end{equation} This first expression $P^0$ gives a lower bound of the
pressure field, and provides a satisfactory approximate
(Fig.~\ref{fig_Pprofile}).
To go further, we use $P^0(x)$ to determine the lowest order velocity profile $v^0_i(z)$
and calculate $\sigma^0_{zz}=2\eta\partial_z v^0_z$.
Eq.~(\ref{Peq}) is then resolved using this calculated numerical
value of $\sigma^0_{zz}$, yielding the next order $P^1(x)$ which in
turn  can be used for further iterations.  Using realistic values for
the parameters corresponding to the  actin cytoskeleton, used in
\cite{FabienGerbal11012000} and \cite{Callan-Jones:2008sf}, the
procedure converges rapidly. Note that in the general case of any
$\alpha$, the first iteration giving $P^0(x)$  has to be performed
numerically, and then the same procedure applies.

This mechanism therefore produces a {\it forward} flow which relies on a pressure build up to $P^*$ in
the gel, here induced by confinement. It requires a
minimal system size $L$ given by taking $x=0$ in Eq.
\ref{Pex}. Interestingly, this can be obtained even for a very low
bare friction coefficient  $\xi_0$, since  the exponential dependence
of the friction on the pressure field permits the effective friction to
reach large values for finite $L$, enabling motion.

On the other hand, Eq. \ref{Pex} shows that $L$ increases faster than linearly with $b$, indicating that this
mechanism would not be significant in the case of a gel on a flat open substrate. In this case, which depicts the
lamellipodium of a cell lying on a flat substrate (see
\cite{Kruse+}), the typical confining length $b$ is
large (of the order of the cell size), and the typical length $L$
necessary to build a strong  pressure gradient is very large ($L>$
cell size). As the pressure gradient in the cell is then much weaker than in the confined case, the friction with
the substrate remains close to its bare value, yielding a much smaller
momentum transfer with the substrate. Additionally, we then expect that the pressure remains below $P^*$, and that no elastic phase is formed, thus preventing forward flow. Without confinement, our model  therefore suggests  a {\it retrograde}  flow, that is in the 
opposite direction to the  moving leading edge, as previously modeled and observed for lamellipodia \cite{Kruse+,Yam:2007ao} on flat substrates.
The flow direction, and therefore the direction of
the pressure gradient  in the gel, constitutes the main difference
between the confinement induced mechanism of motility that we report
here and the standard picture of cells lying on flat substrates.

Experimentally the pressure field can be quantified
indirectly by measuring the effective contact area of the cell membrane using Reflection Interference
Contrast Microscopy (RICM). Fig. \ref{fig_geom}a \cite{note} shows clearly that for a DC confined in a channel a larger contact area at the back of the cell is seen, indicating a backward pressure
gradient in qualitative agreement with our theoretical prediction.

We now argue that this mechanism of confinement induced motility could be used by cells such as DCs to move in confined
environments like channels. 
Extra hypotheses have to be added to the above model in order
to more realistically capture the geometry of  a moving cell. Instead
of an open system, we assume now that the back edge is a thin slice  where the gel is in its elastic regime, which
mimics the uropod  observed at the back
of the cell. Additionally, we assume $\beta P^*\gg 1$ such that the 
friction of the uropod with the channel walls is very large, enforcing $v(0)=0$.
To conserve the total cell mass, we further assume that in the uropod the gel depolymerizes at the speed of the
leading edge $V=v(L)+v_p(L)$ (which defines the over all speed of the
cell). A high depolymerisation rate in the uropod can be justified by the high pressure and a depletion of free actin monomers due to the forward flow. With these hypotheses, the
model presented above mimics a cell moving in a channel with velocity
$V$, and shows that the confinement induced motility mechanism can
indeed be used by cells. Interestingly,  for the parameter values used
in Fig. \ref{fig_Pprofile} the front velocity is calculated to be
$\sim 10\mu\text{m/min}$ which is close to the 
 velocity that can be reached by DCs  in collagen matrices
\cite{Lammermann:2008tf} and in channels (up to $12-15\mu\text{m/min}$) \cite{note}
which significantly is larger than the polymeristation velocity taken as the speed on a flat surface, $4-6 \mu\text{m/min}$ \cite{note,Helden2006}.
\begin{figure}[hbt]
\begin{centering}
\includegraphics[width=0.45\textwidth]{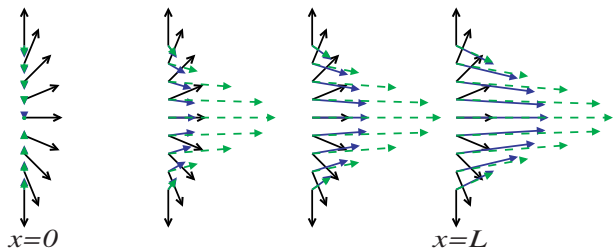}
\caption{\label{fig_vp_profiles}Polarization (black) and flow velocity with (dashed green or gray) and without (blue or dark gray) myosin. Polarization arrows point in the direction of actin polymerisation. The active term is taken as ${\tilde \zeta}(x)\Delta\mu={\tilde \zeta}\Delta\mu (L-x)/L$ for $x>0$ such that there are more myosins at the back of the cell.}
\end{centering}
\end{figure}

Finally, we show that the coupling of the contractile effect of myosins
with the normalized polarization field $p_i$  of actin filaments
(parametrized by its angle $\theta$ with the $x$ axis) can also be
taken into account. The polarization $\theta(z)=\theta_0+\theta_1$ where $\theta_0=-\frac{\pi}{2}(1-\frac{2z}{b})$ is the static configuration which satisfies the normal anchoring
boundary condition and $\theta_1$ is assumed to be linear in $v_p^0$. Taking into account the active stress 
$\sigma_{ij}^{\rm active}={\tilde \zeta}(x)\Delta\mu p_ip_j$ from \cite{Kruse+}, we obtain a perturbative solution around $v_p^0=0$ for the polarization and flow fields (shown in Fig.
\ref{fig_vp_profiles}).
Here it is also useful to consider the
lubrication approximation $b\to 0$. To lowest order in this
approximation, the polarization is given by its static configuration
$\theta_0$ and one obtains a generalized Darcy's law:
\begin{equation}
\label{darcy2}
v(x)= -\frac{b^2}{12\eta}(1+{\tilde \xi}^{-1})\frac{\dd P}{\dd x}  -\frac{b{\tilde \zeta}(x)\Delta\mu }{4\pi\eta},
\end{equation} where ${\tilde \zeta}(x)\Delta\mu$ stands for the
active coupling of myosins to actin filaments (see
\cite{Kruse+} for review), which is to linear order
proportional to the myosin concentration. This equation shows that the
contractile active stress induced by myosins (${\tilde
\zeta}\Delta\mu<0$), increases the velocity of the actin flow, as
shown in  Fig. \ref{fig_vp_profiles}. 

In conclusion, the motility  mechanism of DCs in confined
environments is strikingly different from the standard picture of cell
motility on open flat substrates, and is well captured by our model of
confinement induced motility.  Importantly, this mechanism is widely
independent of adhesion properties with the substrate, since the
mechanism relies on an enhancement of friction due to a pressure
build-up, and does not require specific adhesion proteins. In
particular, this result is compatible with the experiments of
\cite{Lammermann:2008tf}, where it is found that integrin knocked out
DCs are motile only in confined environments.  
Finally, the effect of myosin induced contractility can be
taken into account, and yields a further enhancement of motility, in agreement with experiments \cite{Lammermann:2008tf,note}.


\begin{thebibliography}{27}
\bibitem[{not()}]{note}
\bibinfo{journal}{The experimental protocol will be published elsewhere.}
\end{thebibliography}
\end{document}